\documentclass{svproc}
\usepackage{url}
\usepackage{amsmath}
\usepackage{graphicx}
\usepackage{amssymb}
\usepackage[ruled,vlined]{algorithm2e}
\newcommand{\be}{\begin{equation}}
\newcommand{\ee}{\end{equation}}

\begin{document}
\mainmatter              
\title{IRS Assisted Decentralized Learning for Wideband Spectrum Sensing}
\titlerunning{IRS Assisted DL for Wideband SS}  
%
\author{Sicheng Liu\inst{1} \and Qun Wang\inst{1}
Zhuwei Qin\inst{2} \and Weishan Zhang\inst{3} \and Jingyi Wang\inst{1} \and Xiang Ma\inst{4}}
\authorrunning{Sicheng Liu et al.} 
%
%
\institute{Department of Computer Science, San Francisco State University, San Francisco, CA, 94132
\and
School of Engineering, San Francisco State University, San Francisco, CA, 94132
\and
Department of Electrical and Computer Engineering, George Mason University, Fairfax, VA, 22030
\and
Department of Computer Science, University of Wisconsin-Eau Claire, Eau Claire, WI, 54701
\\
\email{scliu0016@outlook.com, claudqunwang@ieee.org, zwqin@sfsu.edu, \\wzhang23@gmu.edu, Jingyiwang@sfsu.edu, maxiang@uwec.edu}
}

\maketitle              

\begin{abstract}
The increasing demand for reliable connectivity in industrial environments necessitates effective spectrum utilization strategies, especially in the context of shared spectrum bands. 
 However, the dynamic spectrum-sharing mechanisms often lead to significant interference and critical failures, creating a trade-off between spectrum scarcity and under-utilization. 
 This paper addresses these challenges by proposing a novel Intelligent Reflecting Surface (IRS)-assisted spectrum sensing framework integrated with decentralized deep learning. 
 The proposed model overcomes partial observation constraints and minimizes communication overhead while leveraging IRS technology to enhance spectrum sensing accuracy. 
 Through comprehensive simulations, the framework demonstrates its ability to monitor wideband spectrum occupancy effectively, even under challenging signal-to-noise ratio (SNR) conditions.
 This approach offers a scalable and robust solution for spectrum management in next-generation wireless networks.
\end{abstract}

\section{Introduction}

The convergence of the Internet of Things (IoT) and Artificial Intelligence (AI) is poised to revolutionize the way we live and work, enabling advanced applications such as smart manufacturing and autonomous vehicles \cite{genaiqun}. 
By 2025, it is projected that there will be approximately 27 billion artificial IoT (AIoT) connections. Providing reliable connectivity for this massive number of devices is a major driving force behind the development of next-generation wireless networks \cite{iothu}.
To accommodate the rapid expansion of industrial AIoT networks and the growing need for wireless connections, next-generation wireless networks are being deployed over shared spectrum bands to enhance connectivity in industrial applications, such as Citizens Broadband Radio Service (CBRS) 3.5 GHz and 6 GHz band \cite{ssqun}. 

However, employing dynamic spectrum-sharing mechanisms to access new radio frequency bands will increase the potential for substantial interference and critical failures if not properly managed \cite{ss1}. 
Inadequate monitoring of spectrum occupancy can lead to safety issues, unauthorized access to critical systems, and disruptions or damage to industrial processes. 
Therefore, effective spectrum monitoring is essential.
Traditional spectrum monitoring methods, dependent on static models, struggle to adapt to the dynamic and unpredictable nature of modern wireless spectrum usage \cite{sstr1}\cite{sstr2}.

Deep Learning (DL) has emerged as a promising solution for spectrum monitoring by learning spectrum usage patterns from radio frequency measurement data. 
Recently, a DL detector was proposed in \cite{cnnss} to extract the energy correlation features from the covariance matrices and the series of energy-correlation features corresponding to multiple sensing periods. 
A graph learning-based spectrum sensing method was proposed in \cite{gnnss1} that leverages the low-rank property of the received signal strength matrix to improve sensing performance.
Nevertheless, existing DL approaches encounter unique challenges in highly complex and large-scale wireless applications\cite{zhuweifed2}. 
First, distributed spectrum monitoring devices are geographically dispersed and can only observe partial spectrum frequencies or channels relevant to their locations \cite{zhuweiss}. 
Moreover, various impairments in the high-frequency bands,
including shadowing, multipath fading, and path loss over long distances, can significantly increase the false spectrum sensing results and affect the collaborative learning framework’s performance \cite{wz2}. 
To address these challenges, intelligent reflecting surface (IRS) systems can be utilized to mitigate those effects for spectrum activity detection on the sensing device side to enhance monitoring results \cite{irs1}. 
IRS consists of patch antennas printed on a dielectric substrate and controlled by a microcontroller circuit board. This controller adjusts reflection amplitudes and phases of IRS elements for signal control to increase the intended signal strength. 

Thus, this paper proposes a novel on-device learning framework designed to monitor highly uncertain spectrum bands over very wide bandwidths efficiently. 
Our contribution is as follows: 
(1) We proposed an IRS-assisted spectrum sensing model to overcome the pathloss of high-frequency spectrum signals. 
(2) An innovative decentralized deep learning algorithm is proposed to handle partial observations effectively. By minimizing communication overhead and enabling efficient learning from distributed partial observations, the framework aims to enhance spectrum monitoring capabilities. 
(3) Comprehensive simulations have been performed to verify the effectiveness of the proposed algorithms.

The remainder of this paper is organized as follows. Section II introduces the system model and problem formulation. Section III details the proposed Distributed learning-based spectrum sensing detection algorithms. Section IV discusses the simulation setup and results. Finally, Section V concludes the paper.

\begin{figure}
\setlength{\abovedisplayskip}{3pt}
		\setlength{\belowdisplayskip}{3pt}
\includegraphics[width=1\linewidth]{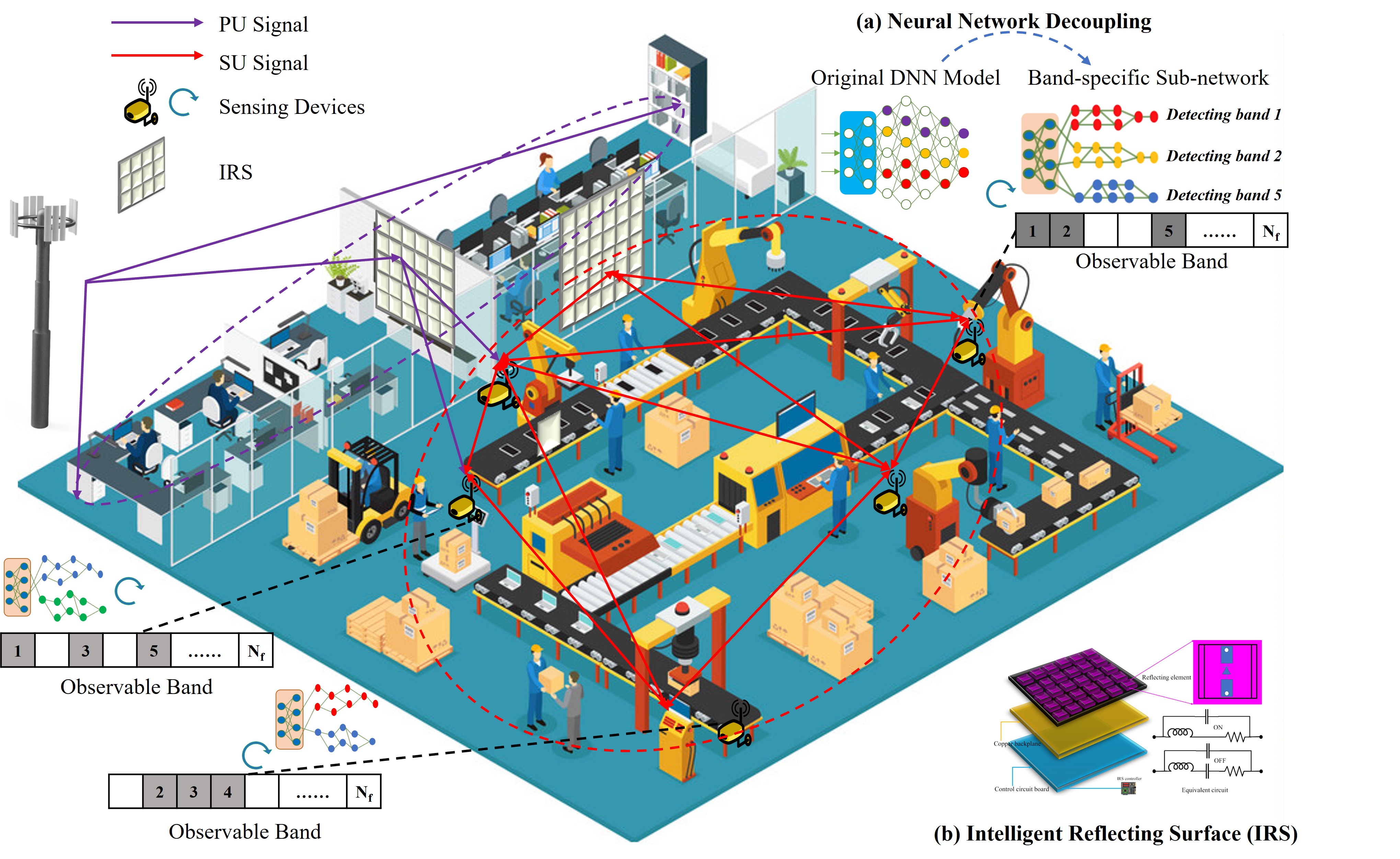} 
\caption{Decentralized Collaborative Spectrum Monitoring with Intelligent Reflecting Surface Enhancement}
\label{sysmodel}
\end{figure}

\section{System Model}
As shown in Fig \ref{sysmodel}, a decentralized collaborative spectrum monitoring system is proposed to monitor radio activities over each band $i$ $(i\in N_f)$ throughout the wideband spectrum pool under partial observations. Several IRSs have been deployed to increase the accuracy of the detection among different sensing devices. 

\subsection{Spectrum Sharing Model}
To measure the transmit activities over different spectrum bands,  the signal model of the power spectrum density (PSD) based measurement data will be first formulated, and the pre-processing steps will be provided to generate and label the training data. 
 {Suppose that a wideband spectrum pool $ B $ is uniformly divided into $ N_f $ bands, where each of them carries a potential spectrum occupancy by a certain primary user (PU). For a CR system with $ I $ Secondary Users (SUs), the time sequence sampled at SU-$ I $, $ \forall I \in [1, I] $, is regarded as collected measurements of received signals \cite{ws2}. Each of these samples can be expressed as a summation of every primary signal reaching SU-$ I $ plus the noise $ z_i $:
}
\begin{equation}
y_i = \sum_{n=1}^{N_f} \tilde{y}_n^i + z_i, 
\end{equation}
 where $ \tilde{y}_n^i $ is the received signal at SU-$ i $ corresponding to the ground-truth $ n $-th primary signal $ y_n $. To extract the spectral features in learning, a pre-processing is adopted by applying the Fourier transform on the autocorrelation of $ y^j $ in (1) at SU-$ j $ \cite{zhuweiss}:

\begin{equation}
y_{\text{PSD}}^i = \text{FT}(\text{Corr}(y^i)), 
\end{equation}
where FT denotes the Fourier transform and Corr(.) computes the signal autocorrelation. Suppose that the dimensionality of $ y_{\text{PSD}}^j $ is $ 1 \times N_w N_f $, where $ N_w $ is related to the spectral resolution of each band inversely. Given all bands with the same bandwidth, the wideband PSD $ y_{\text{PSD}}^j $ can be uniformly segmented into $ N_f $ band-specific PSD vectors of size $ 1 \times N_w $: $ y_{\text{PSD}}^j = [y_{1,\text{PSD}}^j, \ldots, y_{N_f,\text{PSD}}^j] $. To unveil and leverage the inherent correlation of primary signals between different bands, we stack all the band-specific PSD vectors into an $ N_w \times N_f $ matrix as the input training data:
\begin{equation}
Y^i = [y_{1,\text{PSD}}^{jT}, \ldots, y_{N_f,\text{PSD}}^{jT}]. 
\end{equation}

For each channel, the impacts of pathloss, shadowing, white noise and power leakage from neighboring bands will be considered, then the PSD vector for band $n$ will be given as
\begin{equation}
y_{n,\text{PSD}}^i = h_{n}^i x_n +z _{n}^i + \sum_{n' \in \mathcal{B}_n} \eta_{n'} h_{n'}^i x_{n'},\label{eq4}
\end{equation}
where $ h_i x_n $ represents the received PU signal power after considering the channel gain $ h_i$ at $ \text{SU}_i $.
    $ z_n^i $ is the noise PSD at $ \text{SU}_i $ on band $ n $.
  The summation term $ \sum_{n' \in \mathcal{B}_n} \eta_{n'} h_{n'}^i x_{n'}$ accounts for the power leakage from adjacent frequency bands $ n' \in B_n $ into band $ n $.
    $ \mathcal{B}_n$ is the set of all indices of frequency bands adjacent to band $ n $.
   $ \eta_n \in [0, 1] $ is the leakage ratio, indicating the proportion of power from adjacent bands leaking into band $ n $.






\subsection{IRS Enhanced Channel Model}

The channel model integrates the effects of IRS into the traditional path-loss and shadowing model. In this model, the overall channel gain $ h_{n}^i $ is a combination of the direct path gain $ h_{\text{direct}} $ and the reflected path gain $ h_{\text{IRS}} $ via the IRS \cite{irsqun}. 
The direct path gain is given by $ h_{\text{direct}} = \beta \left( \frac{d_0}{d^i_{\text{direct}}} \right)^\alpha 10^{\frac{-\psi^i_{\text{direct}}}{10}} $, where $ d^i_{\text{direct}} $ is the distance between the base station and the user, $ \alpha $ is the path-loss exponent, and $ \psi^i_{\text{direct}} $ represents the shadowing effect. 
IRS introduces additional reflected paths from the PU to the SU. Each IRS element adjusts the phase of the reflected signal to enhance the received power at the SU. The reflected channel gain $h_{\text{IRS},ij}$ is modeled by considering the path from the PU to each IRS element and then reflected to the SU \cite{irs2}.
The reflected path gain through the IRS consisted of two cascade channels from PU to IRS reflecting element $m$ and from IRS to SU, i.e., $h_{i,m}^{P,I}$ and $h_{m,j}^{I,S}$. Where $h_{m}^{P,I}$ is modeled as 
\be
h_{i,m}^{P,I} = \beta \left( \frac{d_0}{d_{im}} \right)^{\alpha}\cdot 10^{-\frac{\psi_{im}}{10}} ,
\ee
where $ d_{im} $ is the distance from PU $ i $ to IRS element $ m $.
$h_{m,j}^{I,S}$ is modeled as 
\be
h_{m,j}^{I,S} = \beta \left( \frac{d_0}{d_{mj}} \right)^{\alpha} \cdot 10^{-\frac{\psi_{mj}}{10}}, 
\ee
where $ d_{mj} $ is the distance from IRS element $ m $ to SU $ j $.

Each IRS element introduces a controllable phase shift $ \theta_m $. The channel gain via an IRS with $ M $ elements can be given as
    \be
    h_{\text{IRS},ij} = \sum_{m=1}^{M} h_{i,m}^{P,I}\cdot e^{j \theta_m} \cdot h_{m,j}^{I,S}.
    \ee
 Total Channel Gain:
    \be
    h_{ij} = h_{\text{direct},ij} + h_{\text{IRS},ij}.
    \ee


\subsection{Problem formulation}
Due to the pathloss fading for the reflecting signal and direction adjustment of the IRS reflecting phase, the power leakage from neighboring bands for the reflecting link will be omitted in this work. Given the input training data from (\ref{eq4}), the CR spectrum sensing problem at SU-$i$ for narrow-band settings can be formulated as a binary hypothesis testing problem either in $H1$ or in $H0$ when band-$n$ is occupied or vacant.
\begin{equation}
\label{eq:H1}
\begin{split}
y_j^n &= w_j^n + h_j^n x_n + \sum_{n' \in \mathcal{B}_n} \eta_{n'} h_j^{n'} x_{n'},~ (H_1\text{: busy})\\
y_j^n &= w_j^n + h_j^n \cdot 0 + \sum_{n' \in \mathcal{B}_n} \eta_{n'} h_j^{n'} x_{n'}, ~ (H_0\text{: idle}).
\end{split}
\end{equation}
where $ y_j^n $ is the received Power Spectral Density (PSD) at $ \text{SU}_j $ on band $ n $. 


For input data $ s_n $ with labels $ \{0, 1\} $ and Deep Neural Network (DNN) parameter set $ W $, the basic single-band spectrum sensing problem can be represented as a function $ f(s_n, W) \in [0, 1] $ \cite{zhuweiss}. In this context, the task of deep learning-based spectrum sensing at a specific $ \text{SU}_j $ on a single band $ n $ is to find the optimal parameter set $ W^* $ that generates the correct hypothesis based on the received PSD $ s_j^n $ of the specific band:

\begin{align}
f(s_j^n | H_1, W^*) &\geq 0.5, \nonumber\\
f(s_j^n | H_0, W^*) &< 0.5.
\end{align}

Due to the representational capability of DNNs, learning-based single-band detectors can be automatically trained with sufficient labeled data, even in the absence of expertise in the underlying signal and channel models. The training objective can be formulated as:
\begin{equation}
\min_W \sum_{ \{ s_j^n, y_n \} \in D } \text{Loss}_b\left( f(s_j^n, W), y_n \right),
\end{equation}
where $ D $ is a dataset of labeled occupancy including PSD and target single bands, using the true occupancy $ y_n = \{0, 1\} $ as labels. $ \text{Loss}_b $ represents the binary cross-entropy loss function based on probability confidence values, defined as \cite{celoss}:
\begin{equation}
\text{Loss}_b(p, q) = - \left[ p \log q + (1 - p) \log (1 - q) \right].
\end{equation}
When extending the single-band case to the multi-band scenario, a key difference aIRSes compared to the conventional cross-entropy loss used in existing classifiers. Specifically, developing a multi-band sensing model aims to distinguish one class out of a total of $ N_c = K^{N_f} $ classes, where $ K $ represents the number of occupancy states per band and $ N_f $ is the number of frequency bands. However, encoding all $ N_c $ occupancy classes of $ N_f $ bands causes the size of the classifier's softmax output layer to grow exponentially with the number of bands.

To address this challenge, we design a novel DNN structure based on multi-class predictors that significantly reduce the model size and computational complexity. By rethinking the output layer design and leveraging distributed learning strategies aided by IRS, our model efficiently captures the multi-band occupancy states without the exponential increase in parameters associated with traditional softmax classifiers.

\section{Distributed Learning-based Spectrum Sensing Design}
To overcome the challenges of partial observations in wideband spectrum sensing, we propose a distributed learning framework that combines decentralized collaborative deep learning with spectrum-specific optimizations. Each device locally trains its model based on observed frequency bands and collaborates via parameter sharing to achieve global learning across distributed sensing devices.

\begin{figure}
\setlength{\abovedisplayskip}{3pt}
		\setlength{\belowdisplayskip}{3pt}
\includegraphics[width=1\linewidth]{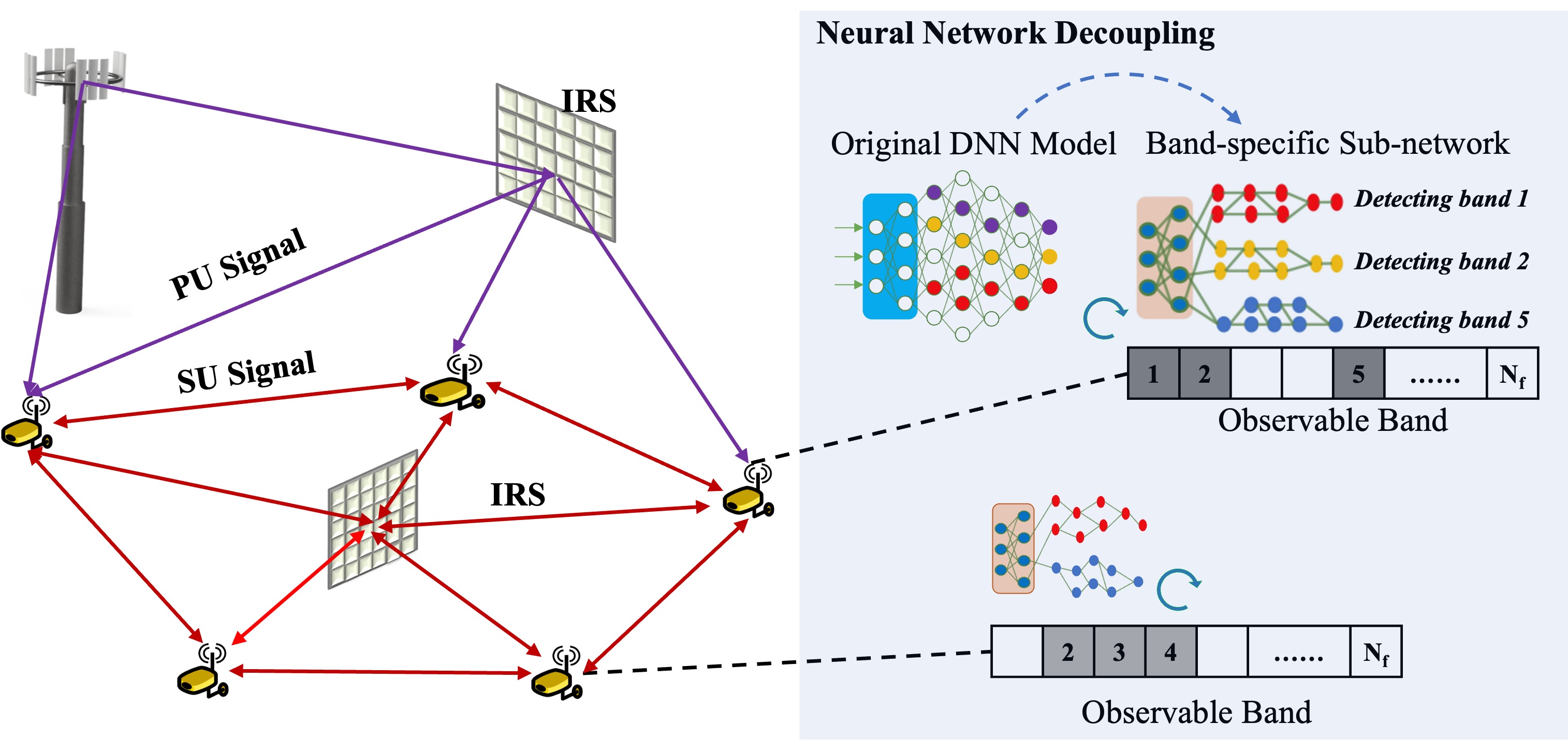} 
\caption{Decentralized Collaborative Neural Network Architecture.}
\label{dcpnn}
\end{figure}

\subsection{Deep Neural Network Architecture}
As shown in Fig. \ref{dcpnn}, the proposed spectrum sensing framework utilizes a multi-task DNN to address the challenges of wideband spectrum sensing under partial observations. The proposed DNN is designed to extract both global and band-specific features from the PSD matrices of observed signals. The architecture is structured hierarchically, integrating shared shallow layers for global feature extraction and band-specific deep layers for localized analysis, optimized for efficient spectrum sensing in distributed environments.

\textbf{1. Shared Shallow Layers:}  
The initial shared layers are tasked with extracting global spectral characteIRStics that are consistent across all frequency bands. The network begins with a sequence of convolutional layers, such as a $3 \times 3$ convolution with 40 filters, followed by batch normalization to stabilize gradients and improve convergence. Rectified Linear Unit (ReLU) activation functions are applied to introduce non-linearity. Max pooling layers, with a pooling size of $4 \times 1$, are utilized to reduce spatial dimensions while preserving critical features. This shared structure ensures that redundant computations are minimized, providing a unified representation of the input PSD matrix \cite{zhuweiss}.

\textbf{2. Band-specific Deep Layers:}  
Following the shared layers, the network adopts specialized sub-networks for each frequency band. These band-specific layers leverage grouped convolutional operations, where the grouping is determined by the number of frequency bands being processed. A special decouple CNN architecture with the third convolutional layer employs $3 \times 3$ kernels with 64 filters and grouped by the number of channels,  ensuring that each group processes features from a specific band \cite{wenshan1}. Batch normalization and ReLU activations are applied consistently across these layers to enhance learning dynamics. The final convolutional layers use average pooling with a kernel size of $4 \times 5$, compressing the feature maps into compact representations that feed into fully connected layers.

Each sub-network contains a fully connected layer designed to handle the band-specific outputs. 
It outputs band-specific predictions using a sigmoid activation function. This structure ensures precise classification of spectral occupancy for individual bands while maintaining computational efficiency.
To handle varying input dimensions, the architecture incorporates adaptive pooling strategies. For instance, Max pooling layers with kernel sizes of $2 \times 2$ are employed in scenarios involving higher channel counts, ensuring consistency in feature extraction across different spectrum configurations. This flexibility ensures the model's robustness across diverse operational settings.



\subsection{Collaborative Learning Framework}
\subsubsection{Local Training}
Each secondary user (SU) trains its local model independently based on its partial observations of the wideband spectrum. Specifically, each SU leverages its observed PSD data to optimize the parameters of the shallow and deep layers of its model. The local training process minimizes a binary cross-entropy loss function for each band-specific sub-network, which is expressed as \cite{xiang1}:
\begin{equation}
\mathcal{L}(W_j) = - \sum_{n=1}^{N_f} \left[ y_n \log f(x_n; W_j)+ (1 - y_n) \log \left( 1 - f(x_n; W_j) \right) \right],
\end{equation}
where $N_f$ is the total number of frequency bands, $x_n$ represents the PSD input for band $n$, $y_n$ is the true occupancy state of band $n$, and $W_j$ denotes the local model parameters of SU-$j$.
This local training step enables each SU to independently capture patterns relevant to its observed spectrum data, while avoiding the need to share raw data, thereby reducing communication overhead \cite{xiang2}.

\subsubsection{Parameter Sharing and Averaging}
To integrate the knowledge from all distributed SUs and achieve global learning, the framework employs a \textit{hierarchical parameter sharing mechanism}, consisting of shallow-layer global averaging and deep-layer band-specific averaging:

 \textbf{(a) Shallow-layer Parameter Sharing:}  
    The parameters of the shallow layers are shared across all SUs to capture general spectral features. This global parameter averaging is performed as:
    \be
    W_s = \frac{1}{J} \sum_{j=1}^{J} W_s^j,\label{shallow}
    \ee
    where $J$ is the total number of SUs, $W_s^j$ represents the shallow-layer parameters of SU-$j$, and $W_s$ denotes the globally averaged shallow-layer parameters.

   \textbf{(b) Deep-layer Parameter Sharing:}  
    The parameters of the deep layers are averaged in a band-specific manner, ensuring that only SUs observing the same band contribute to the parameter updates for that band. This band-wise parameter averaging is expressed as:
    \be
    W_n = \frac{1}{|\mathcal{J}_n|} \sum_{j \in \mathcal{J}_n} W_n^j,\label{deep}
    \ee
    where $\mathcal{J}_n$ represents the set of SUs observing band $n$, $W_n^j$ denotes the deep-layer parameters of SU-$j$ for band $n$, and $W_n$ is the averaged parameter for that band.


\textbf{(c) Cosine Annealing Learning Rate Schedule:} To enhance the convergence properties of our deep neural network training process, we adopt a cosine annealing learning rate strategy\cite{lr1}. This approach gradually decreases the learning rate from an initial value \(\eta_0\) to a minimal rate \(\eta_{\min}\) over a predefined number of iterations \(T_{\max}\). Specifically, at each iteration \(t \in [0, T_{\max}]\), the learning rate \(\eta_t\) is updated according to a half cosine function:

\begin{equation}
\label{eq:cosine_annealing}
\eta_t = \eta_{\min} + \frac{\eta_0 - \eta_{\min}}{2} \left( 1 + \cos\left(\frac{\pi t}{T_{\max}}\right) \right),
\end{equation}
where \(\eta_0\) is the initial learning rate at \(t=0\), and \(\eta_{\min}\) is the lower bound for the learning rate as training nears \(T_{\max}\). 

\subsubsection{Workflow of Collaborative Learning}
The overall training process, including local training and parameter sharing, is summarized in Algorithm~\ref{alg:collaborative_training}, which details the iterative updates for shallow and deep layers:


\begin{algorithm}[h]
\caption{Collaborative training of shallow and deep layers}
\label{alg:collaborative_training}
\KwIn{Dataset $\mathcal{D}$, number of rounds $I$, number of users $J$.}
\KwOut{Optimized parameters $\mathcal{W}$.}

\textbf{Initialize} $\mathcal{W}^j$, $j = 1, \ldots, J$\;
\For{each round $i = 1, 2, \ldots, I$}{
    \For{each $\text{SU}_j$, $j = 1, \ldots, J$ \textbf{in parallel}}{
        $\mathcal{W}^j \leftarrow$ Local training via SGD($i, \mathcal{W}^j, \mathcal{D}$)\;
    }
    \textbf{Parameter averaging:}\;
     \Indp Shallow-layer averaging via Eq.~(\ref{shallow})\;
   \Indp Deep-layer averaging via Eq.~(\ref{deep}) for $n = 1, \ldots, N_f$\;
}
\end{algorithm}

The collaborative learning workflow begins by initializing the local model parameters for each SU. During each training round, local training is performed independently for each SU using its observed data. Parameter sharing then aggregates the knowledge from all SUs through shallow-layer and deep-layer parameter averaging, ensuring effective global learning while accommodating the heterogeneity of SU observations.

\textbf{Remark 1 (Reduced Complexity):} 
By decoupling the dense DNN into smaller band-specific sub-networks and averaging parameters hierarchically, the computational complexity is significantly reduced. Compared to traditional centralized models, the proposed framework minimizes the number of trainable parameters and computational overhead for each sensing device. This makes the system well-suited for resource-constrained environments.

\textbf{Remark 2 (Increased Adaptability):} 
The band-specific averaging mechanism allows the model to dynamically adapt to changing observation conditions. For example, when new sensing devices join the network or the observed spectrum pool changes, the model can seamlessly incorporate the new information without requiring full retraining. This adaptability enhances the robustness of the framework in real-world dynamic spectrum environments.


\section{Simulation Results}
In this section, we will evaluate the performance of collaborative spectrum sensing under partial observations, integrating IRS to enhance the communication environment. The goal is to assess the impact of IRS on sensing accuracy while modeling realistic channel conditions, including path loss, shadow fading, and noise.

The spectrum is divided into $N_f = 20$ frequency bands, each containing $N_w = 64$ frequency points \cite{ws2}. Path Loss Exponent $\alpha =3.71$ and Path Loss Constant $\beta=10^{3.154}$.
 The IRS is strategically placed between the PUs and SUs to provide an additional reflected path for the signals. The IRS consists of $M = 100$ elements, each introducing an adjustable phase shift to optimize the reflected signals. The phases are assumed to be aligned to maximize constructive interference at the SUs.
    Each SU observes a noisy signal matrix of size $(N_w \times N_f)$, where signals from active PUs are scaled by the channel gain and combined with noise.
The IRS enhances sensing accuracy by reflecting PU signals toward the SUs. Signal leakage into adjacent bands is also modeled to simulate real-world spectrum usage.
Each frequency band is labeled as either \textit{occupied} or \textit{vacant} based on the presence of PU signals. Labels are encoded in binary format for machine learning-based spectrum sensing. The simulation is implemented in Python using PyTorch.

\begin{figure}
    \centering
    \includegraphics[width=0.9\linewidth]{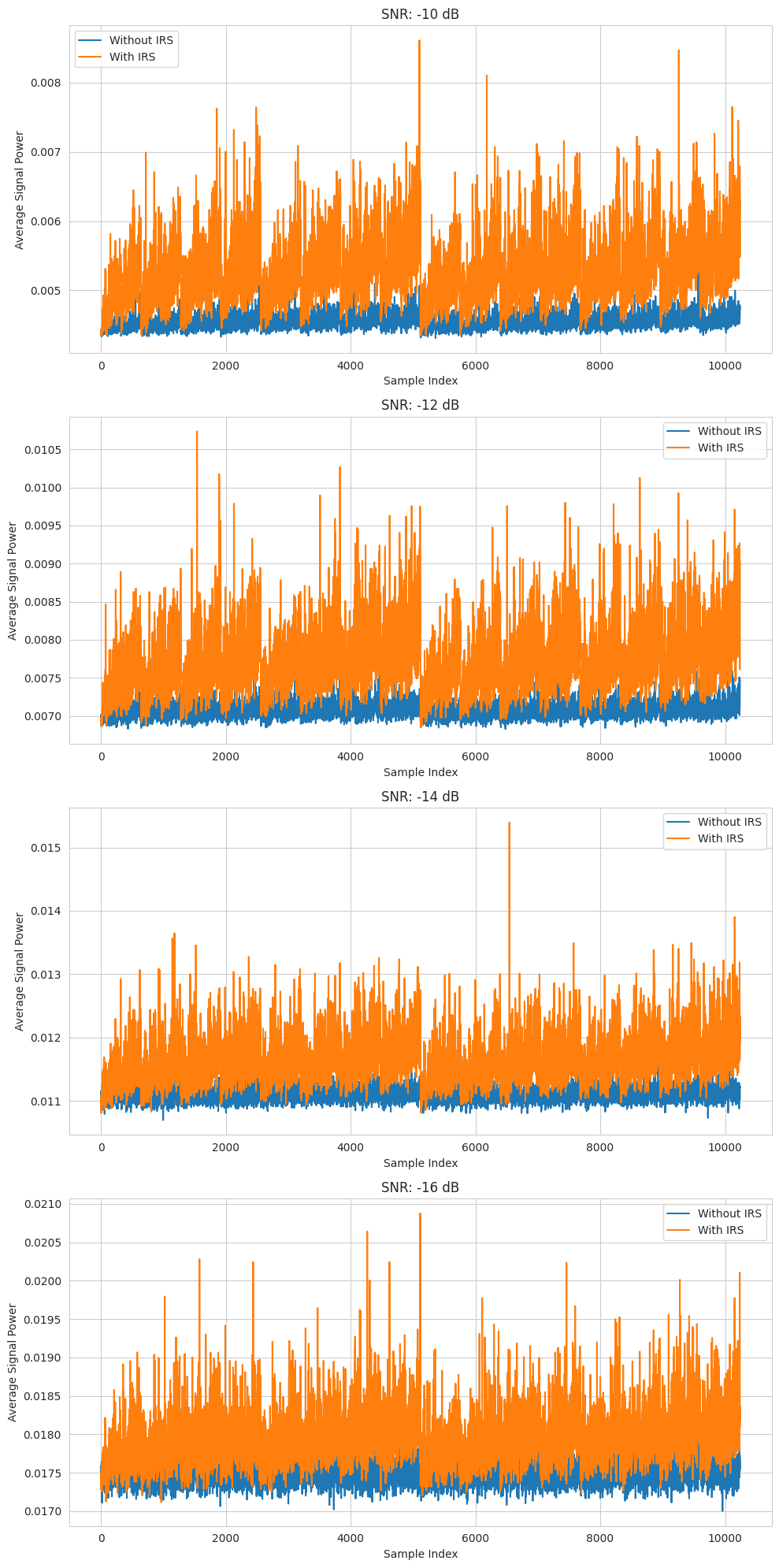}
    \caption{Channel gain with IRS  \label{CSI}}
   
\end{figure}

We first inspect the channel gain brought by the IRS. As shown in Fig. \ref{CSI}, by comparing the received signal strength for scenarios with and without the integration of IRS under different SNR conditions.
The results clearly show that including an IRS can significantly enhance the received signal strength. This is attributed to the IRS's ability to optimize reflected multipath signals, thereby effectively improving the signal power observed by the SUs. As the SNR decreases, the relative gain provided by the IRS diminishes. This is because lower SNR levels correspond to higher noise power, which increasingly dominates the received signal and reduces the relative contribution of the reflected paths.
These findings demonstrate that IRS can effectively augment the communication environment by improving signal power at the receiver.

   

\begin{figure}
    \centering
    \includegraphics[width=0.9\linewidth]{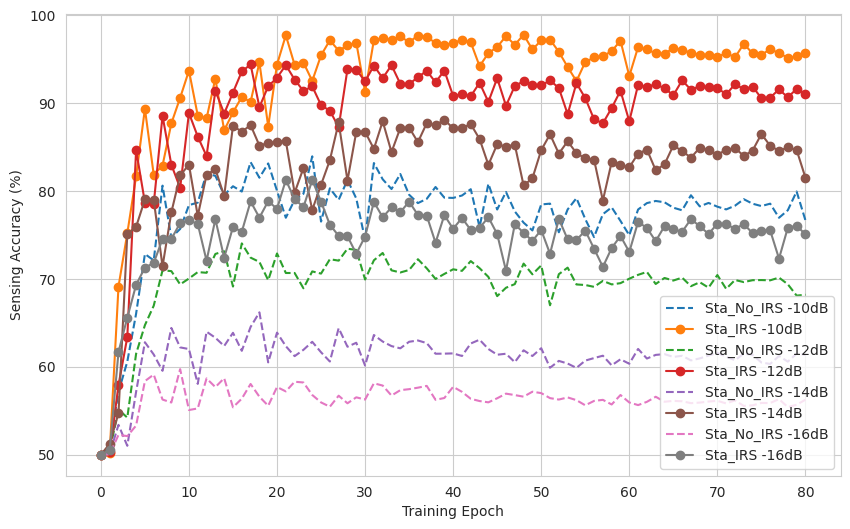}
    \caption{Standalone-based methods performance.   \label{sta1}}
\end{figure}

To validate the performance of the proposed model structure under centralized training, we analyzed the sensing accuracy across multiple SNR conditions, as illustrated in Fig.\ref{sta1}.
The results demonstrate that including an IRS can significantly enhance the model's sensing accuracy across all SNR levels. This is because IRS can improve the signal quality at the receiver, enabling better feature extraction and classification during the training and testing phases.
As the SNR decreases, the overall performance of the model declines due to the increasing dominance of noise in the received signal. This behavior is expected in spectrum sensing tasks as the signal becomes less distinguishable from noise.
However, the proposed model with IRS maintains a consistently higher sensing accuracy compared to the baseline model without IRS. This indicates the robustness of the IRS in enhancing the communication environment even in low-SNR conditions.

At high SNR levels (e.g., -10 dB), the sensing accuracy with IRS approaches near-optimal performance, highlighting the model's ability to leverage the improved channel conditions effectively.
At lower SNR levels (e.g., -16 dB), while the sensing accuracy reduces, the improvement brought by IRS remains significant compared to scenarios without IRS.
These results validate the effectiveness of the IRS in improving the sensing accuracy of the proposed model structure.

\begin{figure}
    \centering
    \includegraphics[width=0.9\linewidth]{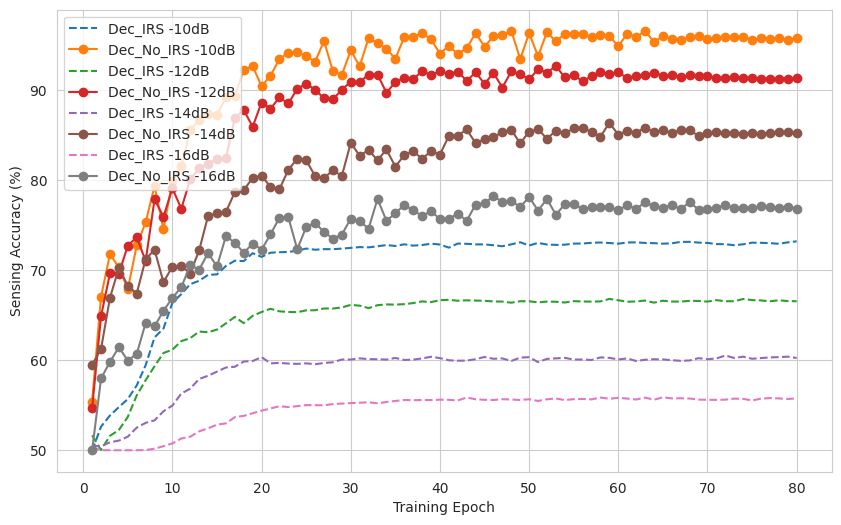}
    \caption{Decopuled method performance\label{decp}}
    
\end{figure}

We further evaluated the proposed distributed algorithm under decentralized training settings as shown in Fig. \ref{decp}. 
Including IRS significantly improves the sensing accuracy across all tested SNR conditions. This demonstrates the effectiveness of the IRS in enhancing communication performance by overcoming limitations imposed by partial observations in decentralized scenarios.
The distributed approach shows faster and more stable convergence compared to centralized training. This is attributed to its ability to better handle the constraints of partial observations, ensuring efficient learning and collaboration among nodes.
The sensing accuracy with the decentralized algorithm stabilizes quickly, even in low-SNR conditions, demonstrating its robustness and reliability in practical applications.
\begin{figure}
    \centering
    \includegraphics[width=0.9\linewidth]{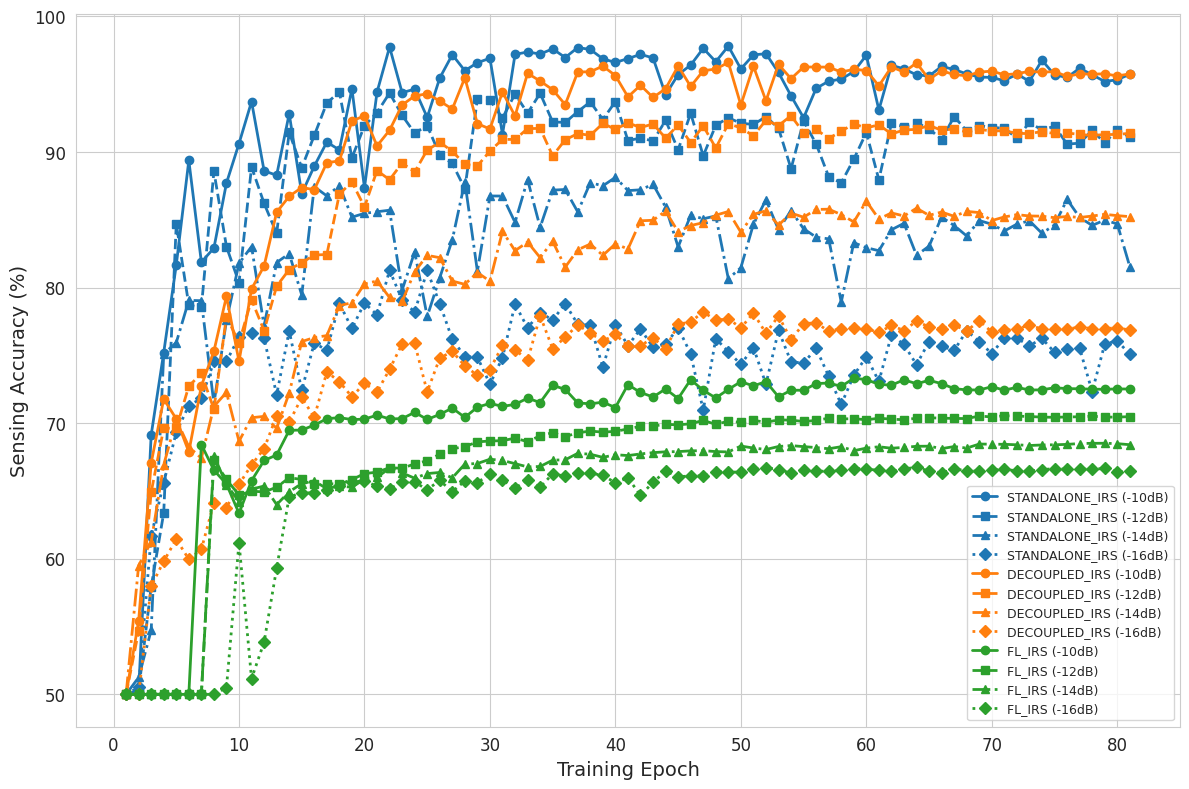}
    \caption{Overall Performance. \label{all}}
\end{figure}

We further compared the performance of the proposed model under centralized and distributed training settings and benchmarked it against traditional federated learning (FL) models \cite{jingyi} \cite{zhuweifed2}. The results are summarized in Fig. \ref{all}.
The proposed model achieves significantly better sensing accuracy than FL models across all SNR conditions. This improvement can be attributed to including personalized and generalized layers, which allow the model to adapt more effectively to local data distributions.
Compared to FL models, the proposed model demonstrates faster convergence during training. This is due to the integration of IRS-assisted enhancements and the effective utilization of local observations through decoupled learning.
The distributed model achieves sensing accuracy comparable to, or even better than, the centralized model. This demonstrates the efficacy of the proposed architecture in leveraging localized observations while maintaining global performance standards.
The personalized and generalized layer structure enables the distributed model to balance local adaptability and global consistency, resulting in superior performance.


\begin{figure}
    \centering
    \includegraphics[width=0.9\linewidth]{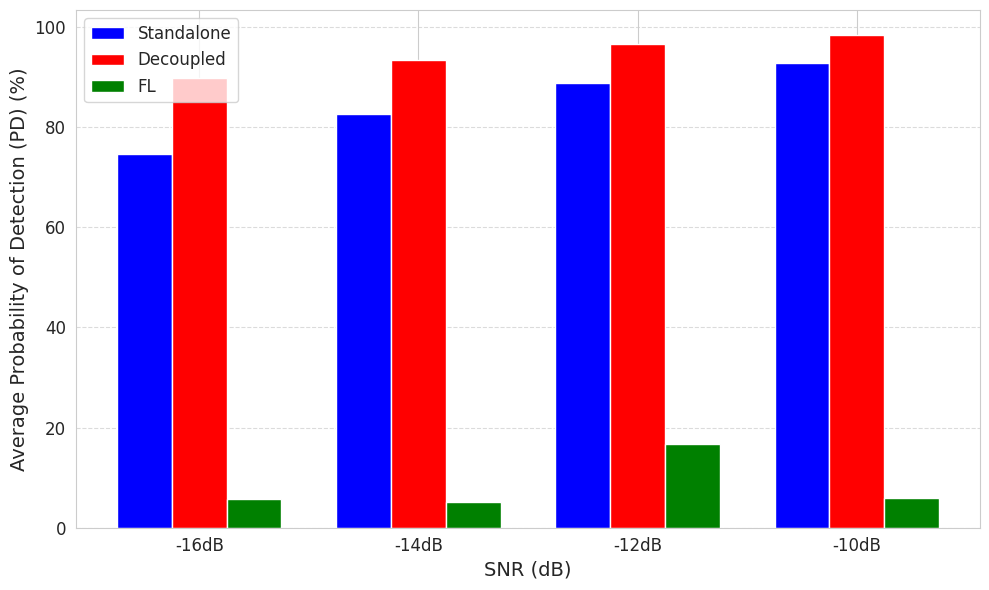}
    \caption{PD Performance. \label{pd}}
\end{figure}
We evaluated the Probability of Detection (PD) performance for three different schemes: Standalone, Decoupled, and Federated Learning (FL). The results are illustrated in Fig.\ref{pd}.
The proposed Decoupled model achieves the highest PD across all tested SNR levels, demonstrating its effectiveness in accurately identifying occupied spectrum bands.
This improvement is attributed to the personalized and generalized layer structure, which balances local adaptability with global knowledge, allowing the model to effectively leverage partial observations.
While the Standalone model performs well, it is consistently outperformed by the Decoupled model due to the latter's ability to share and utilize information across nodes, resulting in better detection accuracy.
The FL scheme exhibits significantly lower PD compared to both the Standalone and Decoupled models. This highlights the limitations of traditional FL in scenarios with partial observations, where the lack of personalized learning results in suboptimal performance.
All schemes show an improvement in PD as the SNR increases. However, the gap between the proposed Decoupled model and the other schemes remains substantial, especially at lower SNR levels (e.g., -16 dB), where the Decoupled model demonstrates robust detection capabilities.
\section{Conclusion}
 An IRS-assisted decentralized deep learning framework is developed in this paper to address the challenges of wideband spectrum sensing under partial observations. By incorporating IRS technology, the proposed model effectively enhances spectrum sensing accuracy by overcoming pathloss limitations and optimizing signal reflection. The hierarchical architecture, combining shared shallow layers and band-specific deep layers, enables efficient feature extraction and reduces computational complexity. Simulation results validate the framework's robustness and efficiency, demonstrating its ability to achieve superior performance in both centralized and distributed settings. Furthermore, the integration of a cosine annealing learning rate strategy accelerates convergence and ensures stability during training. Compared to traditional federated learning approaches, the proposed model exhibits faster convergence, better detection accuracy, and scalability to real-world dynamic spectrum environments. 
	
%
%

\bibliographystyle{IEEEtran}	
	\bibliography{references.bib}








\end{document}